# Conformity Assessments and Post-market Monitoring:
# A Guide to the Role of Auditing in the Proposed European AI Regulation

Jakob Mökander[1], Maria Axente[2], Federico Casolari[3], Luciano Floridi[1,4]

1. Oxford Internet Institute, University of Oxford, 1 St Giles', Oxford OX1 3JS, UK
2. Member of the Advisory Board, UK All Party Parliamentary Group on AI ( APPG AI), London, UK
3. Department of Legal Studies, University of Bologna, via Zamboni 27/29, 40126 Bologna, IT
4. The Alan Turing Institute, The British Library, 2QR, 96 Euston Rd, London NW1 2DB, UK

Correspondence author: Jakob Mokander <jakob.mokander@oii.ox.ac.uk>

**Abstract**

The proposed European Artificial Intelligence Act (AIA) is the first attempt to elaborate a general legal framework for AI carried out by any major global economy. As such, the AIA is likely to become a point of reference in the larger discourse on how AI systems can (and should) be regulated. In this article, we describe and discuss the two primary enforcement mechanisms proposed in the AIA: the conformity assessments that providers of high-risk AI systems are expected to conduct, and the post-market monitoring plans that providers must establish to document the performance of high-risk AI systems throughout their lifetimes. We argue that AIA can be interpreted as a proposal to establish a Europe-wide ecosystem for conducting AI auditing, albeit in other words. Our analysis offers two main contributions. First, by describing the enforcement mechanisms included in the AIA in terminology borrowed from existing literature on AI auditing, we help providers of AI systems understand how they can prove adherence to the requirements set out in the AIA in practice. Second, by examining the AIA from an auditing perspective, we seek to provide transferable lessons from previous research about how to refine further the regulatory approach outlined in the AIA. We conclude by highlighting seven aspects of the AIA where amendments (or simply clarifications) would be helpful. These include, above all, the need to translate vague concepts into verifiable criteria and to strengthen the institutional safeguards concerning conformity assessments based on internal checks.

**Keywords**

Artificial intelligence, Auditing, Certification, Conformity assessment, European Union, Governance, Regulation, Technology.

**This is a pre-print**

The original article is now available online: Conformity Assessments and Post-market Monitoring: A Guide to the Role of Auditing in the Proposed European AI Regulation | SpringerLink

## 1    Introduction

On 21 April 2021, the European Commission published its proposal for a new Artificial Intelligence Act[1] (henceforth AIA). The AIA builds on several recent initiatives and publications that collectively have foreshadowed EU legislation on AI. For example, in the *Ethics Guidelines for Trustworthy AI* (2019), the High-Level Expert Group on AI (AI HLEG[2]) stipulated that AI systems must be ethical, lawful, and technically robust. Building on these guidelines, the European Commission (2020b) subsequently published a *White Paper on AI*, in which the risk-based approach to AI governance that permeates the AIA (more on this in Section 2) was first outlined. Also significantly, the AIA is supposed to constitute a core part of the EU digital single market strategy: indeed, it aims at ensuring the proper functioning of the internal market by setting harmonised rules on the development, placing and use of products and services that make use of AI technologies or are provided as stand-alone AI systems within the Union market.[3] In short, the AIA is a natural continuation of what can be called an EU approach' to AI governance.

The AIA marks a unique milestone that has attracted much attention from policymakers, regulators, commentators, and businesses across the globe. The AIA is the first attempt to elaborate a general legal framework for AI carried out by any major global economy. It is also expected to have a significant impact outside the EU's borders. This impact would be both direct, because the AIA applies to any AI system used in the EU irrespective of where providers are placed (AIA: Article 2); and indirect, because of the 'Brussels effect' (Bradford, 2012, 2020), whereby multinational organisations choose to harmonise all their international practices with EU laws because it is practical to do so.[4] Hence, in addition to constituting proposed legislation in its own right, the AIA is also likely to become a point of reference in the larger discourse on how AI systems can (and should) be regulated.

The initial reactions to the AIA have been many and disparate. Throughout this article, we will return to highlight some of the points raised by different commentators (for a brief overview, see Floridi, 2021). However, the purpose of this article is neither to provide a general commentary on the AIA nor to review, synthesise, or analyse the initial reactions to the AIA from different stakeholders and interest groups. Instead, our concern here is with a more specific yet crucially important question: what is the role of auditing in the proposed EU piece of legislation? Admittedly, the AIA only makes limited, explicit references to auditing. However, in this article, we use the term 'auditing' in a broad sense to refer to structured processes whereby an entity's present or past behaviour is assessed for consistency with relevant principles, standards, regulations or norms (see Section 3). So understood, auditing encapsulates several enforcement mechanisms proposed in the AIA, including the 'conformity assessments' (AIA: Article 43) that providers[5] of high-risk AI systems are expected to conduct and the 'post-market monitoring plans' (AIA: Article 61) that providers must establish to document and analyse the performance of high-risk AI systems throughout their lifetimes. The AIA can thus be interpreted as a proposal to establish a Europe-wide ecosystem for conducting AI auditing, albeit in other words.

---

[1] Its full name reads '*Proposal for a regulation of the European Parliament and the Council laying down harmonised rules on Artificial Intelligence (AIA) and amending certain Union legislative acts'.*
[2] The AI HLEG was an independent expert group set up by the European Commission in 2018. LF was a member.
[3] This explains why the main legal basis for the AIA is Article 114 TFEU, which represents the most relevant legal basis for the establishment and functioning of the EU internal market.
[4] Here, a parallel can be made to the *Global Data Protection Regulation* (GDPR) (The European Parliament, 2016) which, together with the *California Consumer Privacy Act* (2018), has become a de facto global standard for data regulation (Barrett, 2019).

On a few occasions, the AIA does refer explicitly to auditing. However, these references are mostly to be found in the annexes. For example, paragraph 5.3 in ANNEX VII reads as follows: "The notified body shall carry out periodic *audits* to make sure that the provider maintains and applies the quality management system and shall provide the provider with an *audit* report [our italics]". Naturally, sentences like this cannot be understood except as part of the AIA as a whole since it requires an understanding of what exactly is meant by *notified body*, *provider*, and *quality management system* in this context. In this article, we hope to contribute to such a clarification.

Understanding what role audits are expected to play in the proposed EU legislation is essential for practical and theoretical reasons. Most critically, organisations that design and deploy AI systems need clarity on how they can prove adherence to the rules laid out in the AIA. From a practical perspective, the questions thus centre around operational aspects, in particular:

- *material scope*: what is being subject to evaluation?
- *normative baseline*: according to which metrics are AI systems being evaluated?
- *procedural regularity*: what are the roles and responsibilities of different stakeholders throughout the auditing process?

The first goal of this article is to shed light on these operational questions and, thereby, help organisations interpret – and adapt to – the proposed EU legislation on AI.

The second goal is more theoretical. By analysing the role of auditing in the AIA, we seek to anchor the proposed EU legislation in the vast and growing academic literature on AI auditing. We hope that such an analysis will contribute to clarifying what the Commission is proposing. However, it should also be remembered that the AIA is a *proposal*, and as such, it may yet be subject to negotiations and changes.[6] By examining the AIA from an auditing perspective, we seek to provide transferable lessons from previous research about how to refine further the regulatory approach to AI outlined by the Commission.

---

[5] 'Provider' means a natural or legal person, public authority or other body that develops AI systems or that has an AI system that it plans to place on the market, whether for payment or free of charge (AIA: Article 3).

[6] For the GDPR, the process from the first draft to becoming binding took four years.

The remainder of this article proceeds as follows. Section 2 provides a high-level summary of the proposed EU legislation and the societal challenges that it attempts to address. Section 3 reviews previous research to define what we mean by 'enforcement mechanisms' and 'AI auditing' in this context. Section 4 describes and analyses two enforcement mechanisms currently included in the AIA that can also be understood in terms of AI auditing: conformity assessments and post-market monitoring. Section 5 describes the roles and responsibilities assigned to different actors at a corporate, national, and Union level in the AIA. In doing so, it sketches the contours of an emerging European AI auditing ecosystem. Section 6 analyses the scope for non-binding 'ethics-based' auditing within the framework provided by the AIA. The focus is on the codes of conduct to which, according to the AIA, providers of non-high-risk AI systems are encouraged to adhere voluntarily. Section 7 moves beyond what is explicitly proposed in the AIA and provides a gap analysis that identifies areas omitted in the current proposal or where further clarification may help. Finally, Section 8 concludes that, while it constitutes a step in the right direction, the AIA could benefit from incorporating some lessons from previous research on auditing. These include, amongst others, translating vague concepts into verifiable criteria and strengthening the institutional safeguards concerning conformity assessments based on internal checks.

## 2    The Artificial Intelligence Act: a risk-based approach

The proposed EU legislation (i.e. the AIA) represents the most ambitious attempt to regulate AI systems to date (CDEI, 2021a).[7] Most importantly, the AIA seeks to ensure that the AI systems used by – or affecting – people in the EU are safe and respect existing laws and Union values (European Commission, 2021). Importantly, the scope of the AIA also includes the use of AI systems by EU institutions, bodies and agencies (AIA: Recital 12). To prevent risks and harm to public interests and rights that are protected by Union law, the AIA proposes extensive documentation, training, and monitoring requirements on the AI systems that fall under its purview (more on this in Section 3).

However, establishing safeguards against potential harms is not the only objective of the proposed EU legislation. The AIA also stresses that AI systems can support socially and environmentally beneficial outcomes and provide critical competitive advantages to European companies and economies. This claim is well-supported by previous research. For example, AI systems can improve efficiency and consistency in decision-making processes and enable new solutions to complex problems (Taddeo & Floridi, 2018). However, the same elements and techniques that power the socio-economic benefits of AI also bring about new risks for individuals and societies. Specifically, the combination of relative autonomy, adaptability, and interactivity underpins both beneficial and problematic uses of AI systems (Dignum, 2017; Floridi & Sanders, 2004; Russell & Norvig, 2015). Consequently, the capacity to manage the risks posed by AI systems is increasingly becoming a prerequisite for good governance.[8]

---

[7] LF was a member of the CDEI's Advisory Board.

Well aware of this dynamic, the AIA takes as its starting point the twin objectives of promoting the uptake of AI systems and addressing – or at least managing – the challenges associated with such technologies. This 'balanced approach' (AIA: p. 3) has been criticised both by those who contend that the AIA will ultimately stifle innovation (Dechert, 2021), and by those who argue that it leaves Big Tech virtually unscathed and that too little attention is paid to algorithmic fairness (MacCarthy & Propp, 2021). However, we disagree. As we shall argue in the following pages, on the whole, the AIA is a good starting point to ensure that the development of AI in the EU is ethically sound and legally acceptable, as well as environmentally and economically sustainable.

Any attempt to govern AI systems implies making a wide range of design choices. These choices are often difficult and require trade-offs. For example, every regulation needs to define its *material scope* (Schuett, 2019). While there is no commonly accepted definition of AI[9] (Buiten, 2019; Wang, 2019), the definition of AI systems in the AIA[10] is broad by any standard (CDEI, 2021a; Gallo et al., 2021). Hence, the material scope of the AIA is likely to capture decision-making systems that have been in place for decades, in ways that may be problematic. On the one hand, a broad scope of application may prove to be more permanent, since it does not hinge on technical features which are likely to change in the near future. On the other hand, a broad definition risks being over-inclusive, applying to cases that do not need regulation with respect to the regulatory goal, adding unnecessary financial and administrative costs. Such burdens may, in turn, undermine the legitimacy of the regulation. Trying to offset this risk, the AIA proposes different types of obligations for different types of AI systems.

---

[8] The widespread repercussions AI governance has on other policy areas are demonstrated by the fact that the AIA is closely linked to, and coherent with, other initiatives like the *Data Governance Act* (European Commission, 2020a) and the *General Product Safety Directive* (European Parliament/Council, 2001).

[9] According to John McCarthy (2007), AI can be understood as the science and engineering of making intelligent machines. Machine learning (ML), i.e., the study of computer algorithms that can improve automatically through experience and by the use of data (Mitchell, 1997), could thus be viewed as a subset of AI.

[10] For the purpose of the proposed European legislation, the term 'AI system' refers not only to machine learning techniques but also to a wide range of statistical approaches (AIA: ANNEX I).

In fact, the most distinguishing characteristic of the proposed EU legislation is its proportionate, risk-based approach.[11] Simplified, the AIA clusters AI systems into three risk levels:[12] AI systems that pose 'unacceptable risk', 'high risk', and 'little or no risk' (AIA's Explanatory Memorandum: p. 12). The governance requirements differ between the three risk levels. AI systems that are deemed to pose an unacceptable risk, e.g. by posing a clear threat to people's safety, will be straight out banned. This includes the prohibition of AI systems used for general-purpose social scoring and real-time remote biometric identification of natural persons in public spaces for law enforcement (AIA's Explanatory Memorandum: p. 21).

In contrast, AI systems that pose little or no risk are not subject to any interventions stipulated in the AIA, exempt from some specific transparency obligations.[13] According to the AIA, the vast majority of AI systems are expected to fall into this category. However, in between these two extremes, there are a wide range of use cases, or so-called 'high-risk' AI systems, that will be subject to strict obligations before they may be put on the market. To ensure a consistent level of protection from all high-risk AI systems, a common normative standard has been established. That standard is based on the EU Charter of fundamental rights and shall be non-discriminatory and in line with the EU's international trade commitments (AIA: Recital 13). Figure 1 below provides a simplified illustration of the risk-based approach.

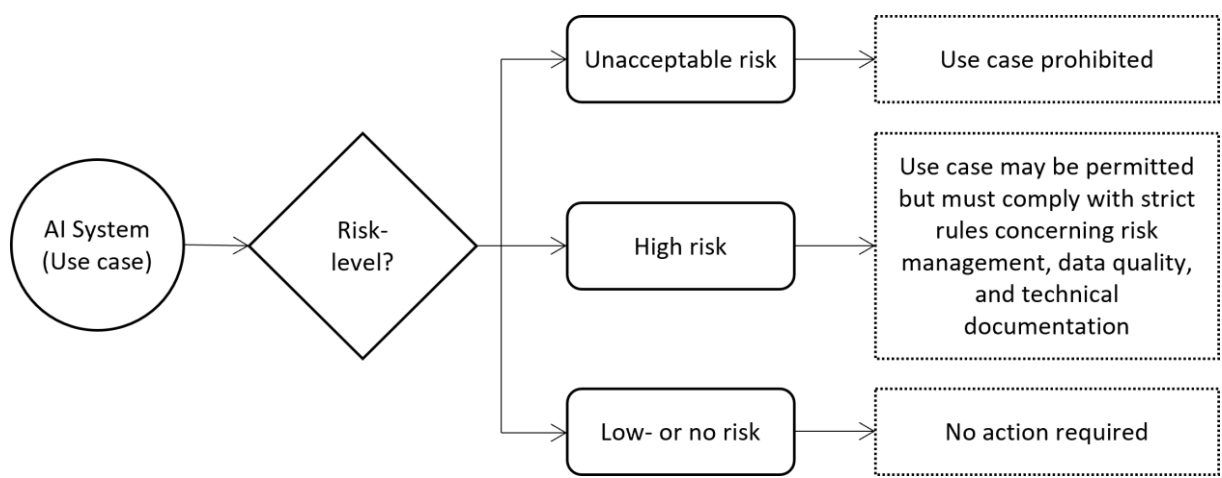

*Figure 1. The risk-based approach to AI governance proposed in the AIA*

---

[11] This risk-based approach can be traced back to publications like the *White Paper on Artificial Intelligence* (European Commission, 2020c) and the *Recommendation of the German Data Ethics Commission* (DEK, 2018).

[12] When determining the risk-level, factors taken into account include the intended purpose of the system, the extent to which the system is likely to be used, and the potential for harm or adverse impacts (AIA: Article 7).

[13] For example, when using a chat bot, users should be made aware of the fact that they are interacting with a machine, rather than a human operator (AIA: Article 52).

The requirements for high-risk AI systems include, but are not limited to, the establishment of a risk management system, the identification and mitigation of known and foreseeable risks, adequate testing and validation (AIA: Chapter 2 of Title III). However, the AIA does not necessarily define rules for specific technologies. Instead, it seeks to establish processes for identifying those use cases requiring additional layers of governance to support specific policy goals. For example, the AIA demands that the technical documentation accompanying a high-risk AI system shall include "a general description of its intended purpose" as well as "a detailed description of the key design choices and assumptions made in the development process" (AIA: ANNEX IV). While such measures contribute to increased procedural regularity and transparency, they also leave significant room for providers to develop and pilot new AI systems.[14]

In Section 4, we will discuss in greater detail discuss how the AIA requirements on high-risk AI systems relate to AI auditing. However, before doing so, it may be helpful to clarify what we mean by AI auditing in this context. Thus, the following section provides a short review of previous research on different enforcement mechanisms for AI governance in general and on AI auditing in particular.

## 3    Previous research: enforcement mechanisms and AI auditing

To be successfully implemented, every regulation needs to be linked to effective enforcement mechanisms, i.e. activities, structures, and controls wielded by various parties to influence and achieve normative ends (Baldwin & Cave, 1999). Responding to the growing need for AI governance, a wide range of enforcement mechanisms have been developed that organisations can employ to ensure that the AI systems they design and deploy are legal, ethical, and technically robust. Some enforcement mechanisms focus on embedding ethical values into AI systems through proactive design (IEEE, 2019). Others are akin to what the CDEI[15] (2021) calls 'assurance techniques'. These include, amongst others, algorithmic impact assessment (ECP, 2018) and certification of AI systems (Scherer, 2016).

The proposed EU legislation on AI includes several enforcement mechanisms. Most notably, the providers of high-risk AI systems that fail to comply with the requirements stipulated in the AIA risk hefty fines. For example, non-compliance with the prohibition of specific uses of AI systems may subject providers to fines of up to 30,000 EUR, or 6% of their total annual turnover, whichever is higher (AIA: Article 71). However, before determining whether a specific AI system is legal, one must consider which mechanisms are available to establish its behaviour and performance (i.e. what it is doing at all). This is where auditing comes in: auditing can be understood as a mechanism that helps organisations verify claims about the AI systems that they design and use.

---

[14] A parallel can be made to what Loi et al. (2020) called *transparency as design publicity*, whereby organisations that design or deploy AI systems are expected to publicise the intentional explanation of the use of a specific system as well as the procedural justification of the decision it takes.

[15] The Centre for Data Ethics and Innovation (CDEI) is an independent advisory body to the UK Government.

Building on previous work (especially Brundage et al., 2020), we define auditing as a structured process whereby an entity's present or past behaviour and performance is assessed for consistency with relevant principles, regulations and norms.[16] Note that while Brundage et al. (2020) focused on *organisational* audits, we stress that the entity in question, i.e. the subject of the audit, can be a person, an organisational unit, or a technical system.[17] Importantly, these different types of audits are not mutually exclusive but rather crucially complementary. To see that this is so, we need only consider the AIA, wherein some legal requirements concern the conduct of organisations that provide AI systems,[18] whereas others concern the technical properties of specific AI systems.[19]

Auditing differs from merely publishing a code of conduct because its primary goal is to show adherence to a predefined baseline (ICO, 2020). So understood, auditing has a long history of promoting trust and transparency in areas like security and financial accounting (LaBrie & Steinke, 2019). Concerning AI governance, auditing can be employed for several distinct yet related purposes. For example, Brown et al. (2021) noted that AI auditing could be used (i) by regulators to assess whether a specific AI system meets legal standards; (ii) by providers or end-users of AI systems to mitigate or control reputational risks; and (iii) by other stakeholders (including customers, investors, and civil rights groups) who want to make informed decisions about the way they engage with specific companies or products. The main takeaway is that all of the above-listed applications of AI auditing align with – and have the potential to support – the stated objectives of the proposed EU legislation.[20]

From an auditing perspective, two enforcement mechanisms included in the AIA are of particular relevance. The first is the conformity assessments that providers need to conduct before putting high-risk AI systems on the market (AIA: Article 43). The second is the post-market monitoring systems that providers shall establish to document and analyse the performance of high-risk AI systems throughout their lifetimes (AIA: Article 61). Let us next consider these in turn.

---

[16] A systems' behaviour and performance cover both 'what' it does and 'how' it does it.

[17] Different stakeholders are accountable for different steps in the process of developing AI systems. As a result, not only the actions of software developers and operators, but also of managers and downstream users, could be subjected to ethics-based audits.

[18] For example, AI providers will be obliged to provide meaningful information about their systems and the conformity assessments carried out on those systems (AIA's Explanatory Memorandum: p. 12).

[19] AI systems should, amongst other properties, be resilient against risks connected to the limitations of the system and against malicious actions that may result in harmful or otherwise undesirable behaviour (AIA's Explanatory Memorandum: p. 30).

[20] In addition to ensuring that AI systems are safe and respect existing laws, the objectives of the AIA include facilitating investments, innovation, and – as already mentioned – the development of a single European market (AIA's Explanatory Memorandum: p. 3).

**4    Conformity assessments and post-market monitoring in the AIA**

In line with the AIA's risk-based approach, high-risk AI systems are only permitted on the EU market if they have been subjected to (and successfully withstood) an ex-ante[21] conformity assessment. Through such conformity assessments, providers can show that their high-risk AI systems comply with the requirements set out in the AIA. Once a high-risk AI system has demonstrated conformity with the AIA – and received a so-called CE marking – it can be deployed in, and move freely within, the internal EU market (AIA: Article 44).

There are three different ways in which these conformity assessments can be conducted. Which type of conformity assessment is appropriate in a specific case depends on the nature of the high-risk AI system. Consider first the many high-risk AI systems used as safety components of consumer products that are already subject to third-party ex-ante conformity assessments under current product safety law. These include, for example, AI systems that are parts of medical devices or toys. In these cases, the requirements set out in the AIA will be "integrated into existing sectoral safety legislation" (AIA's Explanatory Memorandum: p. 4). The reason for this is to avoid duplicating administrative burdens and to maintain clear roles and responsibilities while ensuring a strong consistency among the different strands of EU legislation. However, it also implies that no 'AI specific' conformity assessments will take place. Instead, compliance with the AIA will be assessed through the third-party conformity assessment procedures already established in each sector.

High-risk AI systems that do not fall into the first category are referred to as 'stand-alone' systems. The complete list of stand-alone, high-risk AI systems subject to conformity assessments is found in ANNEX III to the AIA. These include AI systems used in recruitment, determining access to educational institutions, and profiling persons for law enforcement, to mention a few notable examples. All stand-alone high-risk AI systems have to comply with the requirements set out in the AIA. However, providers of stand-alone, high-risk AI systems have two options for how to conduct ex-ante conformity assessments. They can either (a) conduct ex-ante conformity assessments based on internal control, or (b) involve a third-party auditor (i.e. a notified body, more on this in Section 5) to assess their quality management system and technical documentation (AIA: Article 43).

Procedure (a) is only an option where the stand-alone, high-risk AI system is fully compliant with the requirements set out in Chapter 2 of Title III of the AIA. When, in contrast, the compliance is only partial – or harmonised standards do not yet exist – providers are obliged to follow procedure (b). This may seem opaque. However, Figure 2 (below) illustrates through a simple flow-chart when different ways for conducting conformity assessments apply.

---

[21] Ex-ante or 'before the event' conformity assessments take place before a system is placed on the market. In contrast, post-market monitoring is a type of ex-post compliance check.

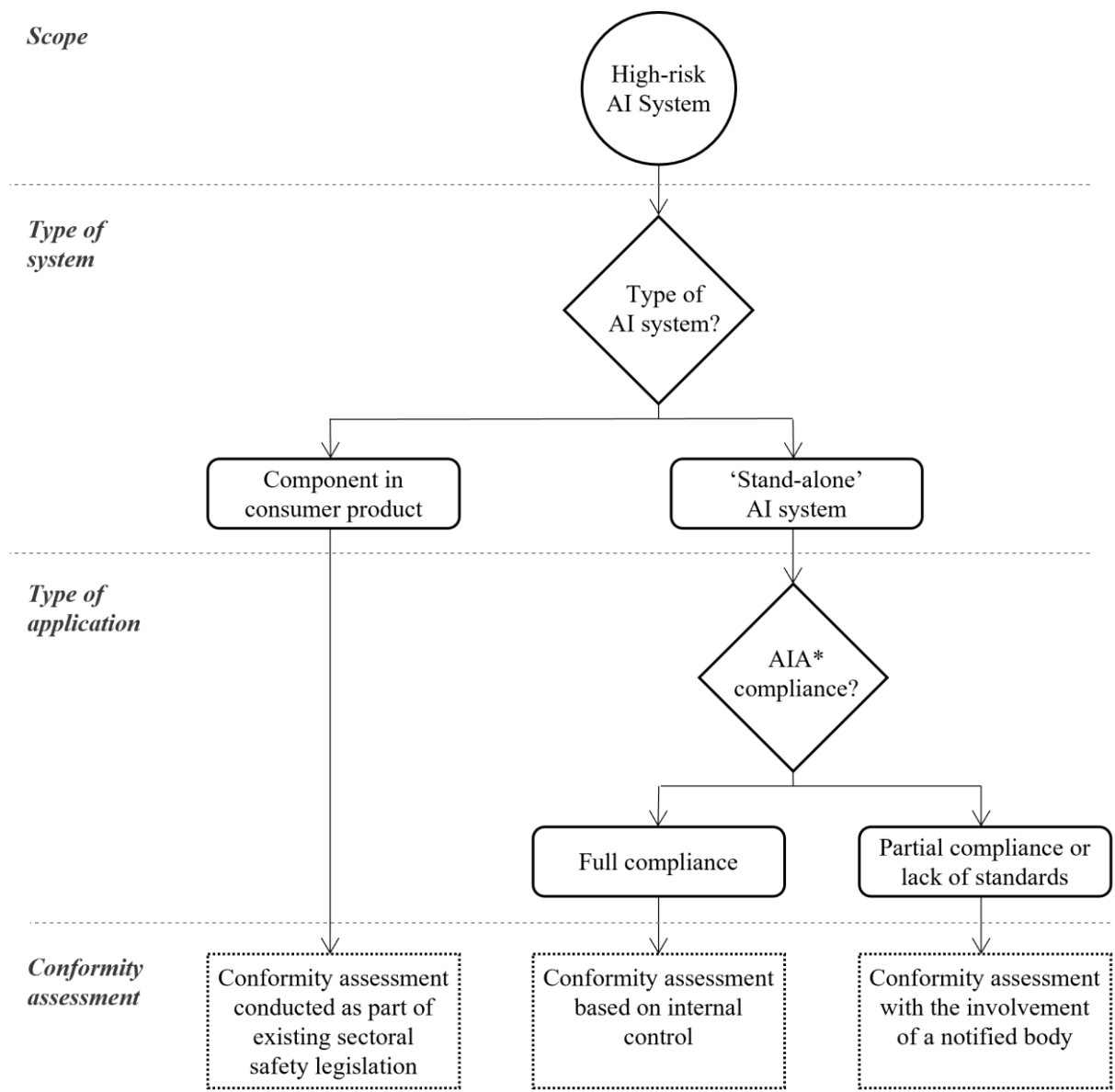

*Figure 2. Ways to conduct conformity assessments for high-risk AI systems*

Ultimately, the legal requirements are the same for all high-risk AI systems. According to ANNEX IV in the AIA, these include, amongst others, obligations on the provider to:

(i)     document the intended purpose of the AI system in question,

(ii)    provide detailed user instructions,

(iii)   disclose the methods used to develop the system, and

(iv)    justify the critical design choices made by the provider.

However, in practice, not all high-risk AI systems will be subjected to third-party (i.e. external) ex-ante conformity assessments. The conformity assessments based on internal control that some providers of stand-alone, high-risk AI systems will have to conduct are more akin to what in the AI auditing literature is referred to as *internal auditing*. These internal checks would include properly documented ex-ante compliance with all requirements of the proposed EU legislation and establishing robust quality and risk management systems per Article 17 in the AIA. In addition, the internal conformity assessment should be accompanied by detailed technical documentation concerning internal governance processes (AIA: Article 18).

Both external and internal audits come with their own sets of strengths and weaknesses. Because external audits help address concerns about the incentives for accuracy in self-reporting (Brundage et al., 2020), they are typically required for formal verification and certification procedures (Yanisky-Ravid & Hallisey, 2018). However, external audits are fundamentally limited by a lack of access to internal processes at the audited organisation (Raji et al., 2020). Hence, they have a limited impact on how AI systems are designed. At the same time, organisations often employ internal audits to check the process in which AI systems are developed (Raji et al., 2020). While they run an increased risk of collusion between auditors and auditee, internal audits can thus constitute a first step towards making informed model design decisions (Saleiro et al., 2018).

Of course, the Commission is well aware of the risks associated with internal audits. However, the AI sector is very innovative, and expertise for AI auditing is only now being developed. Hence, the choice of mechanism design is justified in the AIA by the fact that the providers of stand-alone, high-risk AI systems are best placed to intervene in the early stages of the system development process. Further, while internal conformity assessments rely on the active collaboration of providers of high-risk AI systems, the AIA includes several safeguards against adversarial or negligent behaviour on their parts. For example, after performing the conformity assessment, providers of high-risk AI systems must draw up an EU declaration of conformity[22] (AIA: Article 48). This declaration then becomes part of the officially required documentation accompanying the high-risk AI system, which, in turn, serves as a basis for the CE marking. Here, it should be noted that not only non-compliance but also the failure to act and communicate proactively and transparently can subject providers of high-risk AI systems to penalties. More specifically, Article 71 in the AIA stipulates that the supply of incorrect, incomplete, or misleading information in response to a request from relevant authorities shall be subject to administrative fines.[23]

The outline above provides only a brief sketch of the three different paths through which conformity assessments can be conducted. However, our aim here is only to extract and make visible the information available in the proposed European legislation as currently drafted, not to 'fill in the gaps'. In section 7, we will turn to discuss how the AIA could be amended. Nevertheless, two areas

---

[22] A separate declaration of conformity shall be drawn up for each AI system and kept for 10 years after the AI system has been placed on the market or put into service.

[23] 10,000,000 EUR or, if the provider is a company, 2 % of its total worldwide annual turnover for the preceding financial year, whichever is higher (AIA: Article 71).

where further clarification is needed could already be highlighted at this stage. First, the AIA only provides limited guidance on how – and according to which standards – sector-specific conformity assessments will be conducted in practice. Further, while stressing that that the types of risks posed by an AI system should be evaluated on a sector-by-sector approach (AIA's Explanatory Memorandum: p. 8), the AIA does not provide any sector specific guidance on what type of documentation is needed. Nevertheless, the European Commission stresses that the AIA will be complemented by other, ongoing or planned, initiatives. This includes, for example, revisions of sectoral product legislation such as the *Machinery Directive* and the *General Product Safety Directive* (AIA's Explanatory Memorandum: p. 5).

Second, there is a lack of clarity about which AI systems, precisely, require conformity assessments conducted with the involvement of a third-party (procedure (b) in our typology above). The AIA, as currently drafted, displays a somewhat circular reasoning: procedure (a), i.e., conformity assessments based on internal control is sufficient for stand-alone AI systems that are in compliance with the AIA, but how can providers know if a specific AI system is compliant before the assessment is performed? In the *Commission Staff Working Document* accompanying the AIA,[24] it is suggested that whether or not the conformity assessment needs to follow procedure (b) hinges on the intended use of the AI system in question. For example, the AIA explicitly states that conformity assessments of AI systems intended for remote biometric identification in public spaces will require the involvement of a third-party (Haataja & Bryson, 2021). However, the list of potentially sensitive application areas is likely to expand and change over time, and borderline cases are bound to emerge. Further clarification will thus be needed – both with regards to the criteria used to determine the appropriate conformity assessment procedure for different types of stand-alone AI systems and the safeguards needed to ensure that technology providers don't opt for internal control in cases where the AIA mandates the involvement of third-party auditors.

In addition to the ex-ante conformity assessments described above, providers of high-risk AI systems are also expected to establish and document post-market monitoring systems. The task of post-market monitoring is to document and analyse the behaviour and performance of high-risk AI systems throughout their lifetime (AIA: Article 61). These ex-post assessments are crucially complementary to ex-ante certifications since providers of high-risk AI systems are expected to report any serious incident or any malfunctioning that constitute a breach of Union law (AIA: Article 62). They are also obliged to take immediately any corrective actions needed to bring the AI system under conformity or withdraw it from the market (AIA: Article 21).

---

[24] See the European Commission (2021b) *Commission Staff Working Document : Impact Assessment Accompanying the Proposal for a Regulation of the European Parliament and of the Council*

To detect, report on, and address system failures in effective and systematic ways, providers must first draft post-market monitoring plans that account for, and are proportionate to, the nature of their respective AI systems. The post-market monitoring plan is, in turn, part of the required documentation that constitutes the basis for the conformity declaration (AIA: ANNEX IV). Here, it is important to note that such ongoing, post-market monitoring is intrinsically linked to quality management as a whole. According to the AIA (Article 17), the main objective of the quality management system is to establish procedures for how high-risk AI systems are designed, tested, and verified. However, it should also include procedures for data management, record keeping, and – most importantly for our purposes – procedures for how to implement and maintain post-market monitoring of the high-risk AI system in question.

Legally mandated, post-market monitoring adds a new element and new complexities to corporate quality management systems. Providers of high-risk AI systems are not necessarily the ones operating them. Hence, providers must give users clear instructions on how to use high-risk AI systems and cooperate with them to enable effective post-market monitoring. For example, consider the requirement that high-risk AI systems shall be designed with capabilities to automatically record (or 'log') their operations and decisions (AIA: Article 12). These logs can either be controlled by the user, the provider, or a third party, as per contractual agreements. In any case, however, it is the provider's responsibility to ensure *that,* and plan for *how,* high-risk AI systems automatically generate logs.

The post-market monitoring plan is complementary to the conformity assessment because it is based on a different logic. In the academic literature, a distinction is often made between three complementary yet distinct approaches to AI auditing: *functionality audits* focus on the rationale behind using an AI system; *code audits* entail reviewing the source code of an algorithm; and *impact audits* investigate the types, severity, and prevalence of effects of an AI system's outputs. Whereas the conformity assessments mandated by the AIA entail elements of both functionality audits and code audits, the post-market monitoring plan adds the element of impact auditing. This element is specifically important for AI systems that continue to learn, i.e. update their internal decision-making logic after being deployed at the market.

Combined, the ex-ante conformity assessments and the post-market monitoring mandated by the AIA constitute a coordinated and robust basis for enforcing the proposed EU regulation. However, an enforcement mechanism will only be as good as the institution backing it. Thus, in the next section, we examine the institutional structure proposed in the AIA, i.e. the roles and responsibilities of different stakeholders in ensuring that high-risk AI systems comply with the proposed EU regulations throughout their lifecycles.

## 5    The emergence of an EU AI auditing ecosystem

Ensuring that high-risk AI systems satisfy the various requirements set out in the AIA would require a well-develop auditing ecosystem that consists of two components. First, an institutional structure is needed that clarifies the roles and responsibilities of private companies, national and supranational authorities. This would also include ensuring accountability for different types of system failures. Second, the actors in the ecosystem need access to well-calibrated auditing tools and the necessary expertise to carry out the different steps in demonstrating that high-risk AI systems comply with the AIA. Unfortunately, as noted by the CDEI (2021a), such an ecosystem does not yet exist. Nevertheless, as well shall see in this section, the proposed EU legislation already sketches the contours of an emerging European AI auditing ecosystem.

According to the AIA, the ultimate responsibility to ensure compliance and identify and mitigate potential compliance breaches rests with the providers and users of high-risk AI systems. However, to ensure regulatory oversight, the Commission proposes to set up a governance structure that spans both Union and national levels (AIA 's Explanatory Memorandum: p. 15)[25]. At a Union level, a 'European Artificial Intelligence Board' will be established to collect and share best practices among member states and to issue recommendations on uniform administrative practices (AIA: Article 56). Quite significantly, the AIA does not adopt the 'agencification'[26] approach, which is inherent in the enforcement machinery established in other strands of EU legislation (including the GDPR) In fact, the European Artificial Intelligence Board is not conceived as an independent body having a legal personality. Rather, it is understood as a coordinating structure, chaired by the Commission, where Member States' and Commission's representatives are gathered to facilitate the effective implementation of the AIA.

In addition, the Commission will set up and manage a centralised database for registering stand-alone, high-risk AI systems (AIA: Article 60). The purpose of the database is to increase public transparency and enable ex-post supervision by competent authorities.

At a national level, member states will have to designate a competent national authority to supervise the application and implementation of the AIA. Importantly, this national supervisory authority should not conduct any conformity assessments itself. Instead, it will act as a notifying authority (AIA: Article 59) that assesses, designates, and notifies third-party organisations that, in turn, conduct conformity assessments of providers of high-risk AI systems. In the proposed EU legislation, these third-party organisations are sometimes referred to as 'conformity assessment bodies', but, more often, they are simply called 'notified bodies' (AIA: Article 3.22). To become a notified body, an organisation must apply for notification to the notifying authority of the member state in which they are established.[27]

---

[25] The auditing ecosystem described in the text is subject to some specific adjustments where the AIA interacts with other pieces of EU legislation. This is the case, for instance, of the Union legislation on financial services. According to the AIA, in order to ensure a mutual consistency, the authorities responsible for the supervision and enforcement of the financial services legislation, including the European Central Bank, should be designated as competent authorities (AIA: Article 63.4). Moreover, where Union institutions, agencies and bodies fall within the scope of the AIA, the European Data Protections Supervisor shall act as market surveillance authority (AIA: Article 63.6).

[26] The expression 'agencification' is normally used to refer to the proliferation of EU agencies within the EU legal order which has gained a terrific momentum from the 1990s: (Chamon, 2016).

The main task of a notified body is to assess and approve the quality management systems that providers of high-risk AI systems use for the design, development, and testing (AIA: ANNEX VII). Further, the notified body shall examine the technical documentation for each high-risk AI system produced under the same quality management system. Based on these assessments, the notified body shall then determine whether both the quality management system and the technical documentation satisfy the requirements set out in the AIA. Where conformity has been established, the notified body shall issue an EU technical documentation assessment certificate.[28] Figure 3 below provides an overview of the relationship between different private organisations and institutional bodies in the process of assessing and certifying stand-alone, high-risk AI systems.

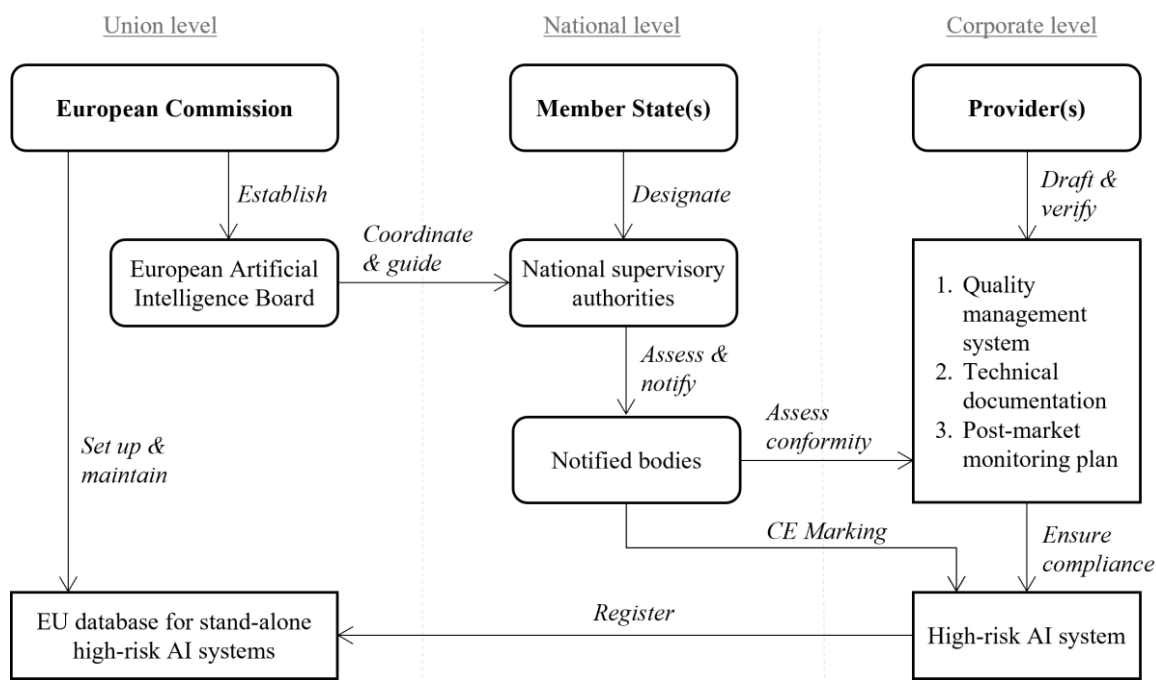

*Figure 3. Roles and responsibilities during conformity assessments with the involvement of third-party auditors*

---

[27] Conformity assessment bodies established in third countries with which the Union has an agreement may be authorised to carry out the activities of notified bodies under this Regulation (AIA: Article 39).
[28] Note that this procedure is only applicable for the conformity assessments of stand-alone high-risk AI systems that require the involvement of third-party auditors (see Figure 2 on page 10 for guidance).

Admittedly, Figure 3 (above) gives a somewhat idealised picture of the roles and responsibilities outlined in the AIA. First of all, the relationships – here indicated by directional arrows – are in reality bidirectional. For example, although the national supervisory authority is responsible for assessing and notifying conformity assessment bodies, it does so based on the application and material submitted by organisations that wish to be notified.[29] After notification, each notified body is then assigned a unique identification number by the Commission (AIA: Article 35). Similarly, while the notified body is responsible for carrying out conformity assessments, providers of high-risk AI systems have an obligation to make the relationship work. That includes collaborating with the notified body, providing the notified body and the national surveillance authority with timely access to all resources and documents that are necessary for a comprehensive assessment to take place,[30] and reporting any severe incidents or malfunctioning of their high-risk AI systems directly to the national surveillance authority.[31] To delivering on these expectations, and comply with the reporting structure proposed in the AIA, providers and users of AI systems may need to appoint new roles within their organisations.

There is also a second sense in which Figure 3 is a simplification. It makes the process looks clear and solidified. In reality, the proposed EU legislation is quite vague and leaves significant room for interpretation: the language used in the AIA is highly technical, and, in several instances, multiple terms are used to refer to the same concept. For example, in the AIA, the terms 'notified body' and 'conformity assessment body' seems to be used interchangeably (AIA: Article 3.21 and 3.22). However, based on the tasks ascribed to the notified bodies, they could also have been called 'auditing bodies'.

Similarly, what the AIA calls' notifying body' is equivalent to what is commonly known as 'accreditation body'. Most EU member states already have national accreditation bodies, and the AIA (Article 30) even highlights that these can be designated as notifying authorities. In Section 7, we shall discuss different points that demand clarification in greater detail. Before doing so, however, the next section will explore the potential scope for soft governance within the AIA.

## 6    The scope for soft governance within the AIA

In this section, we argue that the proposed European legislation should be seen as a complement to – and reinforcement of – the wide range of initiatives launched by both regulators and technology providers in recent years to ensure that AI systems are legal, ethical and technically robust. Specifically, we stress that there will remain a demand for voluntary 'ethics-based' audits (or other types of assurance) that allow organisations to validate claims about their AI systems, and that go over and above demonstrating compliance with the AIA.

---

[29] To become notified, conformity assessment bodies must demonstrate that they have the structure, competence, and resources required to fulfil their tasks.

[30] When public authorities and notified bodies need to be given access to confidential information or source code to examine compliance, they are placed under binding confidentiality obligations.

[31] Such notification shall be made as soon as possible and, in any event, no later than 15 days after the provider becomes aware of the serious incident or the malfunctioning (AIA: Article 62).

To do so, however, we must first take a step back and consider the distinction between soft and hard enforcement mechanisms. Hard governance refers to systems of rules elaborated and enforced through institutions to govern agents' behaviour (Floridi, 2018). Examples of hard enforcement mechanisms range from legal restrictions on system outputs to the prohibition of AI systems for specific applications (Koene et al., 2019). So understood, both the conformity assessments and the mandatory post-market monitoring procedures discussed in the previous section fall into the category of hard enforcement mechanisms. In contract, soft governance embodies mechanisms that exhibit some degree of contextual flexibility, like subsidies and taxes. Put differently, while hard governance refers to legally binding obligations, soft governance includes non-binding guidelines (Erdelyi & Goldsmith, 2018).

Hard and soft enforcement mechanisms often complement and mutually reinforce each other (Hodges, 2015). In the case of AI governance, this is especially true since laws may not always be up to speed in sectors that experience fast-paced technological innovation. Further, decisions made by AI systems may deserve scrutiny even when they are not illegal. Hence, there is always room for post-compliance, ethics-based auditing (Mökander et al., 2021), whereby organisations can prove adherence to voluntary standards that go over and above existing regulations.

In and of itself, the proposed EU legislation constitute hard governance. However, the AIA also leaves room for soft governance in general and post-compliance, ethics-based auditing in particular. Most notably, providers of non-high-risk AI systems are encouraged to draw up and apply voluntary codes of conduct (AIA: Article 69) related to their internal procedures and the technical characteristics of the systems they design and deploy. The critical difference between these voluntary codes of conduct and the other requirements in the AIA is that they focus on *process management* rather than *goal management*. This leaves individual organisations free to either draw up guiding principles of their own, adopt guidelines recommended by the European Artificial Intelligence Board, or declare adherence to any other set of standards relevant for their specific industry or use case.

In the context of the AIA, the European Commission has at least two reasons for encouraging the voluntary use of codes of conduct. The first is to foster the voluntary application of the requirements set out in the AIA, even to use cases not subjected to mandatory conformity assessments. It should be noted that – as of now – the proposed EU legislation imposes no restrictions or obligations on AI systems that are deemed to constitute little or no risk, such as AI-enabled video games and spam filters (O'Donoghue et al., 2021). However, depending on their technical specifications and intended purpose, such systems may also benefit from compliance with the requirements set out in Chapter 2 of the AIA concerning data quality, traceability, technical robustness, and accuracy.

The second objective is to promote post-compliance ethical behaviour. Even providers of high-risk AI systems may benefit from adopting voluntary codes of conduct[32] that go over and above the requirements set out in the AIA. Therefore, providers have good reasons to subject themselves to ethics-based auditing: just as organisations seek to certify that their operations are sustainable from an environmental point of view (IEEE, 2019), or demonstrate to consumers that products are healthy through detailed nutritional labels (Holland et al., 2018), the documentation and communication of the steps taken to ensure that AI systems are ethically-sound can play a positive role in both marketing and public relations (Mökander & Floridi, 2021). For example, by contributing to procedural regularity in, and transparent communication about, how AI systems are designed and deployed, ethics-based audits can help organisations manage financial and legal risks (Koene et al., 2019), improve public relations (EIU, 2020), and gain competitive advantages (European Commission, 2019).

As our analysis in this section has demonstrated, the European Commission encourages the voluntary adoption of codes of conduct and supports the emergence of complementary, soft governance mechanisms that sit on top of the AIA. This is promising. However, the AIA does not provide guidance on *whether* and *how* adherence to voluntary codes of conduct will be assessed. This is a missed opportunity, since there is in fact much regulators that can do to support the feasibility and effectiveness of ethics-based auditing procedures that emerge bottom up from the assurance needs of different stakeholders in the AI ecosystem (Mökander & Axente, 2021). In the next section, we discuss this omission alongside other areas where further guidance may be required.

## 7    The need for further guidance

The overall strategy to implement the AIA is clear. Nevertheless, in several areas, further guidance is needed. This section highlights seven such areas with direct implications for the effectiveness and feasibility of auditing AI.

I.    <u>Level of abstraction</u>.[33] As many commentators have already noticed, some expectations in the AIA seem 'too idealistic' and will thus require 'a lot more guidance' (Gallo et al., 2021). Consider the data quality requirement that training, validation, and testing data sets shall be relevant, representative, free of errors, and complete (AIA: Article 10.3). While this is a laudable vision statement at a high level of abstraction, it may not be feasible to expect data sets to be completely 'free of errors' in practice. Setting the bar too high, or articulating requirements in too abstract terms, can backfire since rules that cannot be translated into operational terms are likely to be regarded only mechanically as a box-ticking exercise. Moreover, unrealistic expectations may undermine the legitimacy of the framework as a whole (Power, 1997).[34] Again, setting high-level expectations is an important exercise in its own right, since expectations help shape the behaviour of different actors in multi-agent ecosystems (Minkkinen et al., 2021). Nevertheless, to be feasible and effective in practice, the AIA needs to provide further guidance on lower, and more detailed, levels of abstraction. That is, high-level visions for data management and software development need to be broken down into applicable industry standards and evaluation metrics.[35] This is particularly important from an auditing perspective, since audits presuppose a realistic benchmark against which to audit.

---

[32] Such codes of conduct may, for example, concern commitments to environmental sustainability, stakeholders' participation in the design of AI systems, or the diversity of development teams.

[33] A level of abstraction (LoA) is a finite but non-empty set of observables, which are expected to be the building blocks in a theory characterised by their very choice (Floridi, 2008). Different LoAs can be nested, disjoined, or overlapping and need not be hierarchically related (Floridi, 2017). However, this is not a relativist approach: a question is always asked for a purpose, and different LoAs can 'fit' the purpose more or less successfully.

II. <u>Material scope</u>. The material scope of the proposed EU legislation is opaque. In some cases it looks very broad. For instance, the definition provided in ANNEX I to the AIA encapsulates several software-developing techniques, including machine learning approaches such as deep neural networks, logic- and knowledge-based approaches like expert systems, as well as statistical approaches such as Bayesian estimation and search and optimisation methods. The idea that a single regulatory approach could be designed in such a way as to tackle the issues associated with each one of these technologies is problematic. Ultimately, all that these technologies have in common is that they process data.[36] A more narrowly defined scope may help providers of AI systems, third-party auditors, and national authorities direct their resources more effectively. Also problematic is the decision to include an exhaustive list of high-risk AI systems in the proposed legislation. On the one hand, as stressed by the European Data Protection Board (EDPB) and the European Data Protection Supervisor (EDPS) in a Joint Opinion on the AIA (EDPB/EDPS, 2021) adopted on 18 June 2021, this technique might create a 'black-and-white effect', undermining the risk-based approach of the Proposal. On the other hand, the list misses some types of uses which are likely to involve significant risks, e.g., the use of AI for military applications, for determining insurance premiums, or for health research purposes.[37] Problems related to the material scope of the AIA may also come from the decision to exclude explicitly the international law enforcement cooperation (AIA: Article 2.4). Taken together, both the broad definition of 'artificial intelligence' and the attempt to exhaustively list applications that fall within the material scope of the AIA risk undermining the very purpose of the regulation. Given that the AIA is justified with reference to the challenges associated with the complexity, degree of unpredictability, and partial autonomy of specific AI systems (AIA's Explanatory Memorandum: p. 2), further clarification is required as to how the material scope (as currently defined), is linked to that regulatory goal.

---

[34] Audits can be viewed as rituals of verification that build trust through procedural regularity (Power, 1997). Hence, it is essential for the legitimacy of the process that the standard outcome is positive.

[35] While standards play an important role in any coordinated response to the risks posed by AI systems (Cihon, 2019), they are of particular importance for audits, since these presuppose a sound baseline to audit against.

[36] But so do human decision-makers. And, since they also make mistakes and produce discriminatory or inconsistent outcomes (Kahneman, 2011), the use of AI systems can sometimes lead to more objective and potentially fairer decisions (Lepri et al., 2018).

III. <u>Conceptual precision</u>. At times, as also stressed in the above-mentioned EDBP/EDPS Joint Opinion, the language used in the AIA is vague and imprecise. This is particulary the case with regard to the concept of 'risk to fundamental rights', which should also be aligned with the EU data protection legislation. Likewise, the vague terminology used in AIA Article 5 to identify the prohibited uses of AI runs the risk of making such limitations meaningless in practice. In fact, pursuant to that article AI systems that "deploy subliminal techniques beyond a person's consciousness to distort a person's behaviour in ways that may cause harm" are prohibited. However, digital mediation inevitably influences human users, for example, by nudging an individual's preferences through positive reinforcement or indirect suggestion (Thaler & Sunstein, 2008; Yeung, 2017). Hence, further guidance is needed regarding which kinds of distortions[38] the AIA refers to as prohibited. Also vague are the cases mentioned in the AIA which may lead to a derogation from the conformity assessment procedure. According to Article 47.1, derogations are possible "for exceptional reasons of public security or the protection of life and health of persons, environmental protection and the protection of key and infrastructural assets". Yet, it is possible to shed some light on that provision by referring to the case law of the European Court of Justice interpreting the classical derogations from the law of EU internal market. Further guidance could be provided by the GDPR's implementation, with particular regard to the unprecedented challenges posed by the reaction to COVID-19 pandemic.[39] That said, the fact remains the Article's wording appears too broad. A more precise (and narrower) definition of cases where derogations may be invoked under the AIA would be advisable. Another example where clarification is needed concerns the duty providers have to report serious incidents to responsible national authorities. Article 62 in the AIA states that such notifications shall be made immediately after the provider has established a causal link between the AI system and the incident. However, it is often difficult to establish causal links, especially for externalities that occur due to indirect chains of events that evolve over time (Dafoe, 2017).[40] A last point that should be better clarified is related to the control of conformity of AI systems already in use. Pursuant to Article 83.2 of the Proposal, those systems should be excluded from the scope of the Regulation, unless they are subject to "significant changes in their design or intended purposes". The provision does not offer further details thereon and the related threshold remains unclear. Some additional elements contributing to clarify the wording of Article 83.2 may be inferred from Recital 66 of the Proposal, which specifies that conformity re-assessment shall take place "whenever a change occurs which may affect the compliance". Even though that threshold is related to AI systems which were already subject to a conformity assessment, it could be also applied to pre-existing AI systems. A more accurate definition of the situations covered by Article 83 would be in any case necessary. In short, the AIA would benefit from further guidance on how vague concepts like 'subliminal distortion techniques' or 'causal links' should be interpreted in practice.

---

[37] In fact, since AI technologies are quickly evolving, there is a real risk that any detailed list of so-called high-risk use-cases will be obsolete by the time the AIA comes into effect.

[38] In a recent article in Nature, Köbis et al. (2021) distinguished between four main roles through which both humans and machines can influence ethical behaviour. These are role model, advisor, partner, and delegate. It is, in particular, AI agents acting as enablers of unethical behaviour (partners or delegates) that may let people reap unethical benefits while feeling good about themselves, a potentially perilous interaction.

[39] A clear illustration is given in this respect by the debate concerning the use of mobile applications to combat and exit from the COVID-19 crisis. Such a debate has led to the development of a supranational toolbox to ensure the respect of privacy and data protection. See European Commission (2020b) and Bradford et al. (2020).

IV. <u>Procedural guidance</u>. While the logic behind the conformity assessments and the post-market monitoring activities mandated in the AIA is clear, many details concerning how these should be conducted in practice have yet to be spelt out. For example, Article 20 in the AIA stipulates that the logs shall be kept for a period that is "appropriate in the light of the intended purpose of the high-risk AI system". However, the AIA does neither say how long is appropriate, nor does it suggest who is responsible for determining this (e.g. the provider, a notified body, national authorities, or the Commission). Similarly, in ANNEX VII to the AIA, it is stated that notified bodies shall carry out periodic audits to make sure that the provider maintains and applies the quality management system following the technical documentation provided during the conformity assessment. However, the AIA does not specify how often periodic audits should be conducted or how such audits are triggered.[41] Finally, the AIA focuses exclusively on AI systems aimed for the market.[42] However, in applied settings, the distinction between basic and market-oriented research is not always clear – and even AI systems used for internal purposes may pose ethical risks. Taken together, these examples highlight a need for further procedural guidance on how conformity assessments and post-market monitoring should be conducted in practice.

---

[40] One option may be to include elements of observation-based impact auditing in the post-market monitoring plan, for example with the help of so-called 'oversight programs' (Etzioni & Etzioni, 2016) that monitor and evaluate system outputs continuously. But the AIA makes does not mention any such options.

[41] Article 44 in the AIA states that certificates shall only be valid for the period they indicate, which shall not exceed five years. However, it is unclear whether the periodic audits mentioned in ANNEX VII refer to the re-assessments that are required to extend the validity of a certificate for further periods, or to periodic audits during the continuous operation of an already certified AI system.

V.  Institutional mandate. The European Commission and the national supervisory authorities, supported by The European Artificial Intelligence Board, have mandates to implement the hard enforcement mechanisms proposed in the AIA.[43] However, while the Commission's powers are clearly identified in the Proposal, the role and mandate of the European Artificial Intelligence Boards remain unclear.[44] The decision to exclude the independence of the Board, which is subject to a significant (if not excessive) control by the Commission, is also debatable. Strong criticisms concerning that institutional solution may be found in the EDPB/EDPS Joint Opinion on the AIA, where the two entities stress the need to recognize more autonomy to the Board through a clearer identification of its nature and powers. On a different note, the AIA does not prevent national authorities from keeping specific regulatory prerogatives in implementing the relevant obligations.[45] Such a situation risks reproducing in the domain at stake the same fragmented approach emerging from the national implementation of GDPR (European Parliament, 2021). Moreover, the AIA does not include any institutional safeguards to maintain the integrity of the voluntary codes of conduct that it encourages providers of non-high-risk AI systems to adopt. This is problematic (EDPB/ESPS, 2021) since the adoption of voluntary codes of conduct can be undermined by unethical behaviours like 'ethics shopping', i.e. mixing and matching ethical principles from different sources to justify some pre-existing behaviour, or 'ethics bluewashing', i.e. an organisation making unsubstantiated claims about AI systems to appear more ethical than one is (Floridi, 2019). Hence, any set of ethical principles will only be as good as the public institution backing it (Boddington, 2017). A potential solution would be to create or designate an independent entity that authorises organisations that, in turn, conduct ethics-based audits to check whether providers of non-high-risk AI systems adhere to their stated codes of conduct.[46] Given that the AIA already sketches the contours of a Europe-wide AI auditing ecosystem, one opportunity would be to leverage the same institutional structure to provide assurance also for post-compliance, ethics-based audits.

---

[42] For example, while AI systems intended to distort human behaviour are prohibited, the European Commission explicitly states that research for legitimate purposes should not be stifled by the prohibition (AIA: p 18).
[43] Except in cases where the AIA interacts with specific sectoral policies of the Union.
[44] The European Artificial Intelligence Board should be "responsible for a number of advisory tasks" (AIA: p. 35). However, the AIA does not specify *how,* and with which *mandate*, the Board will operate in practice.
[45] AIA: Recital 71, recognizing the Member States' right to elaborate artificial intelligence regulatory sandboxes.
[46] Note that an obligation to demonstrate adherence to officially communicated codes of conduct is compatible with the voluntary nature of the code of conduct itself.

VI. <u>Resolving tensions</u>. When designing and operating AI systems, tensions may arise between different ethical principles for which there are no fixed solutions (AI HLEG, 2019). For example, a particular ADMS may improve the overall accuracy of decisions but discriminate against specific subgroups in the population (Whittlestone et al., 2019a). Similarly, different definitions of fairness – like individual fairness, demographic parity, and equality of opportunity – are mutually exclusive (Kusner et al., 2017; Friedler et al., 2016). Given these unresolved normative tensions, it is encouraging that the conformity assessments proposed in the AIA focus on making implicit design choices visible through the disclosure of technical documentation. Organisations are, and should be, free to strike justifiable ethical trade-offs within the limits of legal permissibility and operational viability. However, organisations that develop AI systems respond to various stakeholders who often have divergent interests. European regulators could help providers of AI systems understand and account for these diverse sets of interests, e.g. by complementing the requirements set out in the AIA with further guidance on how to resolve tensions between conflicting values, such as accuracy and privacy, as well as on how to prioritise between conflicting definitions of normative concepts, like fairness, in different situations.

VII. <u>Checks and balances.</u> Although high-risk AI systems are subject to conformity assessments, the enforcement of the requirements set out in the AIA is less stringent than it appears (MacCarthy & Propp, 2021). This is primarily because – for most high-risk AI systems – the conformity assessments will be based on internal checks conducted by the system provider itself. Similarly, while providers must draw up an EU declaration of conformity and give a copy of it to the relevant national authorities upon request (AIA: Article 48), *how* providers ensure compliance with the AIA is not disclosed to the public. This lack of checks and balances may be problematic because pursuing rapid technological progress leaves little time to ensure that AI systems are robust and ethical (Whittlestone et al., 2019b). Moreover, there is always a risk of adversarial behaviour during conformity assessments.[47] Thus, companies find themselves wedged between the benefits of innovation and social responsibility and may not act ethically in the absence of oversight (Turner Lee, 2018). Fortunately, there are several ways of strengthening the conformity assessment process outlined in the AIA. One way would be to impose even stricter transparency obligations so that the conformity assessment process – including the trade-offs made in designing a specific high-risk AI system – are disclosed to the wider public. Another option would be to subject the quality management system put in place by individual providers of high-risk AI systems to ad-hoc audits by independent third parties. Some guidance could also be provided by the ECJ's case-law. However, if we consider the balancing exercise showed so far by the Luxembourg judges in the digital domain, it appears evident that their contribution will be far from decisive. Not only is the relevant case law fragmented, which prevents the emergence of a unitary approach to be replicated in the different strands of EU legislation (Fontanelli, 2016), but it also does not remove the need for private parties to engage in a delicate and unpredictable balancing act.[48]

---

[47] An example of such behaviour was the diesel emission scandal, during which Volkswagen intentionally bypassed regulations by installing software that manipulated exhaust gases during tests (Conrad, 2018).

## 8   Conclusions

We hope to have shown that, on the whole, the proposed EU legislation is a good starting point for balancing the prospective benefits from promoting responsible innovation and providing proportionate safeguards against the risks posed by AI systems. In particular, the risk-based approach outlined in the AIA is promising insofar as it shifts the focus from technology to policy. Going forward, this means that it will be less important to label a specific technical system 'AI' and more important to scrutinise the normative ends for which the system is employed.

Further, the enforcement mechanisms proposed in the AIA (specifically the conformity assessments and the post-market monitoring) bridge a critical gap. Hitherto, providers of AI systems have been encouraged to adopt and adhere to voluntary ethics principles (Hagendorff, 2020; Jobin et al., 2019). However, central questions – such as according to which metrics AI systems should be evaluated and who should be accountable for different system failures – have remained unanswered (Floridi & Cowls, 2019). By outlining hard enforcement mechanisms and proposing an institutional structure with the mandate to implement these, the AIA provides a framework for preventing, reporting on, and allocating accountability for different kinds of system failures.

However, despite these merits, the proposed EU legislation still leaves some room for improvement. In this article, we have argued that the AIA de facto sketches a EU-wide ecosystem for auditing AI systems, albeit in other words. We believe it should make this explicit and plan it strategically. Conformity assessments based on internal checks, for example, are akin to what in the AI auditing literature is called *internal audits*; conformity assessments based on technical documentation with the involvement of a notified body resemble what is known as *external audits*; and the post-market monitoring that providers of high-risk AI systems will have to conduct follows the same methodological logic as *continuous auditing*. It may be preferable to move even further in the direction of 'conformity assessment', avoid any reluctance, and commit to fully supporting an EU-wide auditing ecosystem that is able to provide both compliance and post-compliance levels of assurance.

---

[48] See for instance Case C-507/17 *Google LLC* EU:C:2019:772 and (Susi, 2019).

Of course, there may be good reasons for choosing different terminology and we are aware of the fact that the language used in the AIA echoes the solutions adopted in other pieces of the EU legislation, starting from the legislation concerning the market surveillance and compliance of products.[49] However, by anchoring the AIA in the existing literature on AI auditing, valuable lessons can be learned from previous research. For example, auditing presupposes a predefined baseline to audit against. Hence, vague concepts like 'distorting behaviours' or 'causal links' must be translated into practically verifiable criteria for providers of AI systems to demonstrate adherence to the AIA. Similarly, the risks associated with internal audits are well known. Hence, the AIA would benefit from the inclusion of additional institutional safeguards concerning the enforcement of conformity assessments based on internal control.

By discussing the limitations and omissions of the current draft of the AIA, this article does not seek to diminish the many and great merits of the proposed EU legislation. On the contrary, we support the approach adopted. That is why, in this article, we have aimed to highlight areas where potential amendments to the AIA would help strengthen its overall effectiveness in contributing to good AI governance in the EU and beyond.

---

[49] See Regulation (EU) 2019/1020 with which the AIA presents strong interactions.